\documentclass[sigconf, nonacm]{acmart}

\usepackage{alltt}
\usepackage{relsize}
\usepackage{booktabs}
\usepackage{mdframed}
\usepackage{color}
\usepackage{boxedminipage}
\usepackage{fancyvrb}

\usepackage{url}
\usepackage{fancybox}%
\usepackage{placeins}
\usepackage{adjustbox}
 \usepackage{verbatim}
\usepackage{lscape}

\definecolor{mygray}{rgb}{0.7,0.7,0.7}

{\begin{list}{}%
         {\setlength{\leftmargin}{#1}}%
         \item[]%
}
{\end{list}}

\newcommand{\smalltt}[1]{\ifmmode{\mbox{\smaller\texttt{#1}}}\else{\smaller\tt #1}\fi}

{ \medskip
  \noindent
  \let\emph=\textbf
  \begin{boxedminipage}{\columnwidth}\em}%
{ \end{boxedminipage}}

\newdimen\qdx
\newdimen\qda
\newdimen\qdb
\def\rrrr#1#2#3#4{\newdimen\qd\qd=#4 %
\qdx=\qd\multiply\qdx by 5\divide\qdx by 4
\qda=\qd
\qdb=\qd
\multiply\qda by #1\divide\qda by #3\multiply\qdb by #2\divide\qdb by #3\advance\qdb by -\qda
    \leavevmode\hbox to \qdx{\hfil\vbox{%
    \hbox{\vrule\vbox{\hrule\hbox to 1\qd
            {\vrule depth0pt height0.7ex width \qda\color{mygray}%
 \vrule depth0pt height0.7ex width \qdb\hfill}\hrule}\vrule}
    }\hfil}}

  \renewenvironment{thebibliography}[1]{%
    \begin{oldthebibliography}{#1}%
       \setlength{\parskip}{0.2ex}%
       \setlength{\itemsep}{0.2ex}%
  }%
  {%
    \end{oldthebibliography}%
  }

\usepackage{balance}
\usepackage{makecell}

\usepackage[english]{babel}
\usepackage{graphicx}
\usepackage[shortlabels,inline]{enumitem}
\usepackage{enumitem}
\usepackage{threeparttable}

\newcommand\vldbavailabilityurl{https://github.com/mbrotos/lock-pred}
 
\newcommand{\dbtable}[1]{\texttt{#1}}

\begin{document}
\title{Lock Prediction for Zero-Downtime Database Encryption }

\author{Mohamed Sami Rakha}
\affiliation{%
  \institution{Toronto Metropolitan University}
  \city{Toronto}
  \state{ON} 
  \country{Canada}
}
\email{rakha@torontomu.ca}

\author{Adam Sorrenti}
\affiliation{%
  \institution{Toronto Metropolitan University}
  \city{Toronto}
  \state{ON} 
  \country{Canada}
}
\email{adam.sorrenti@torontomu.ca}

\author{Greg Stager} 
\affiliation{ 
  \institution{IBM Canada Lab}
  \city{Toronto} 
  \state{ON} 
  \country{Canada}
}
\email{gstager@ca.ibm.com}

\author{Walid Rjaibi} 
\affiliation{ 
  \institution{IBM Canada Lab}
  \city{Toronto} 
  \state{ON} 
  \country{Canada}
}
\email{wrjaibi@ca.ibm.com}

\author{Andriy Miranskyy}
\orcid{0000-0002-7747-9043}
\affiliation{%
  \institution{Toronto Metropolitan University}
  \city{Toronto}
  \state{ON} 
  \country{Canada}
}
\email{avm@torontomu.ca}

\def\RQOne{\textbf{How can we predict the next table locked?}}

\def\RQTwo{\textbf{How can we predict the next data page locked?}}

\def\RQThree{\textbf{How does the lock prediction changes?}}

\begin{abstract}

Modern enterprise database systems face significant challenges in balancing data security and performance. Ensuring robust encryption for sensitive information is critical for systems' compliance with security standards.  Although holistic database encryption provides strong protection, existing database systems often require a complete backup and restore cycle, resulting in prolonged downtime and increased storage usage. This makes it difficult to implement online encryption techniques in high-throughput environments without disrupting critical operations.

To address this challenge, we envision a solution that enables online database encryption aligned with system activity, eliminating the need for downtime, storage overhead, or full-database reprocessing. Central to this vision is the ability to predict which parts of the database will be accessed next, allowing encryption to be applied online. As a step towards this solution, this study proposes a predictive approach that leverages deep learning models to forecast database lock sequences, using IBM Db2 as the database system under study. In this study, we collected a specialized dataset from TPC-C benchmark workloads, leveraging lock event logs for model training and evaluation. We applied deep learning architectures, such as Transformer and LSTM, to evaluate models for various table-level and page-level lock predictions. We benchmark the accuracy of the trained models versus a Naive Baseline across different prediction horizons and timelines.

The study experiments demonstrate that the proposed deep learning-based models achieve notable improvements, with average accuracy reaching up to 49\% for table-level lock prediction and 66\% for page-level prediction, outperforming the Naive Baseline. By anticipating which tables and pages will be locked next, the proposed approach is a step toward online encryption, offering a practical path toward secure, low-overhead database systems.

\end{abstract}

\maketitle

\ifdefempty{\vldbavailabilityurl}{}{
\vspace{.3cm}
\begingroup\small\noindent\raggedright\textbf{Artifact Availability:}\\
The source code, data, and/or other artifacts have been made available at \url{\vldbavailabilityurl}.
\endgroup
}

\section{Introduction}

Efficient database management is essential for ensuring high performance transaction processing, particularly in enterprise-scale database management systems (DBMS) such as IBM Db2~\cite{IBMDb2Link}. One of the most pressing challenges for modern organizations is ensuring robust data security through encryption. As cyber threats become more sophisticated and data privacy regulations become stricter, encrypting databases has become an essential requirement for organizations handling sensitive information~\cite{bertino2005database}. The urgency is further amplified by the impending threat of quantum computing, which has the potential to break widely used encryption schemes~\cite{zhang2020quantum, zhang2023making}.

Database encryption can be applied at different levels, including file system encryption, tablespace encryption, and column-level encryption~\cite{shmueli2014implementing}. However, these approaches often require trade-offs between security, storage efficiency, and performance. File-system encryption, for instance, does not fully secure data stored at rest within a database, while column-level encryption, though highly granular, can degrade performance due to computational overhead. The most effective approach to balancing security and efficiency is holistic database encryption, in which the entire database is encrypted at once. This method ensures comprehensive protection without requiring significant application changes~\cite{mattsson2005database}.  

However, holistic database encryption presents a major challenge: online encryption or continuous re-encryption of existing databases while maintaining zero downtime and avoiding substantial storage overhead~\cite{rjaibi2018holistic, ibm_patent}. Currently, enabling online encryption on an operational database often requires a costly and time-consuming backup and sometimes a full restore process, making the database temporarily unavailable for queries~\cite{greenwald2013oracle}. The primary technical challenge is to online encrypt or re-encrypt the entire database without disrupting active transactions or requiring additional storage resources.

To address this challenge, we envision a solution for online database encryption that applies encryption in sync with actual system activity, avoiding the need for full database downtime or added storage overhead. A key component of this solution is the ability to anticipate which parts of the database will be accessed and locked next, allowing encryption tasks to be scheduled just in time. As a step towards this solution, this study explores a deep learning-based approach for predicting database lock sequences in IBM Db2. The ability to forecast which tables or data pages will be locked next may enable proactive encryption strategies, ensuring data security while minimizing performance degradation~\cite{zhang2017workload, fotache2023framework, OrcaleOnlineTime}. Specifically, we propose a Transformer-based  model~\cite{nassiri2023transformer} and an LSTM-based model~\cite{sherstinsky2020fundamentals} to predict (1) the next table to be locked 
 and (2) the next data page to be locked. To evaluate the predictive capabilities of these models, we conduct experiments across multiple forecasting horizons, specifically assessing their performance in predicting sequences of the next 2, 3, and 4 locks. This multi-horizon analysis provides insights into the models' scalability and effectiveness in capturing longer-term dependencies in lock sequences.  We generalize predictions across multiple lock types (e.g., shared, exclusive, intent)~\cite{IBMDB2Modes}.
 
For benchmarking purposes, we compare the performance of our deep learning models against a Naive Baseline that assumes the next lock will be identical to the previous one. 
The lock prediction is a step toward achieving high-performance database management systems that proactively mitigate encryption-induced overheads such as downtime and data storage. In this study, we address the following research questions:

\begin{enumerate}[label=\textbf{RQ\arabic*:}]
	\item {\RQOne} For table-level lock prediction, the deep learning models outperform the Naive Baseline across all horizons, capturing complex locking patterns more effectively. 
    
    \item {\RQTwo} For page-level lock prediction, the global deep learning models demonstrate higher accuracy and greater stability. Surprisingly, the Naive Baseline remains highly competitive when compared to local deep models, often matching or surpassing their performance in certain scenarios.  
     \item {\RQThree} The results show that lock prediction accuracy remains stable over time, with no significant degradation across different forecasting horizons.  

\end{enumerate}

This study contributes to the growing body of work on intelligent database management, with a step towards online encryption with zero downtime and storage overhead. In addition to online encryption, the findings of this research have significant implications for database performance optimization, proactive concurrency control, and adaptive encryption scheduling. By integrating predictive lock management into Db2, enterprises can achieve workload-aware caching, dynamic deadlock avoidance, and encryption-aware access control~---~advancing the vision of self-tuning, security-enhanced databases~\cite{thomasian2024heterogeneous}. The contributions of this study can be summarized as follows. 
\begin{itemize}[leftmargin=*]
\item  \textbf{Deep Learning for Lock Prediction: } We explore the performance of deep learning models to predict database lock sequences. We benchmark the reported accuracy against a simple Naive Baseline for lock predictions.

\item  \textbf{Support Different Lock Levels: } We build prediction models for different lock levels (table-level and page-level), ensuring a step toward online encryption while maintaining performance efficiency.

\item  \textbf{Empirical Insights: } We validate our approach on real-world Db2 traces and TPC-C~\cite{tpcc_v511} benchmark workload, demonstrating the models' accuracies in predicting table-level and page-level locks. %

\item \textbf{Temporal Evolution of Lock Patterns:} We analyze how lock prediction accuracy and patterns change over time, providing insights into how lock predictions react to workload dynamics.

\end{itemize}

The rest of the paper is organized as follows. Section 2 reviews related work on database lock prediction, workload modelling, and buffer pool optimization. Section 3 outlines the methodology, including data collection, feature extraction, and model selection. Section 4 presents the experimental results and evaluates the performance of the proposed predictive models. Section 5 discusses the implications of the findings, while Section 6 addresses key threats to validity, limitations, and directions for future research. Finally, Section 7 concludes the paper by summarizing the main contributions.

\begin{figure*}[t]
    \vspace*{-6mm}
    \includegraphics[width=1.0\textwidth]{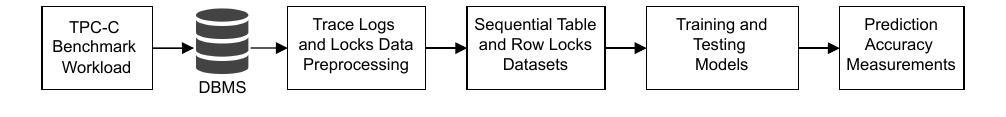}
    \caption{Overview of the lock prediction pipeline.}
    \label{fig:Overview}
\end{figure*}

\section{Related Work}

The prediction of data access patterns has been a subject of interest in database management, particularly in optimizing concurrency control, buffer management, and security mechanisms~\cite{dan1995characterization, elnaffar2004intelligent, duggan2011performance, ma2018query, shi2021hierarchical}.  Recent advancements in machine learning have enabled more sophisticated approaches to forecasting lock behavior, facilitating proactive concurrency control and workload-aware resource allocation. Several studies have investigated data access prediction and resource management in relational database systems, leveraging machine learning techniques to enhance performance, mitigate contention, and strengthen security mechanisms~\cite{rjaibi2018holistic,martin2006workload, chen2021revisiting, nanda2020comprehensive,OrcaleOnlineTime}. This section reviews relevant research in the domains of database lock prediction, workload modeling, and adaptive buffer pool optimization.

\subsection{Lock Contention and Concurrency Control}
Traditional concurrency control mechanisms such as two-phase locking (2PL) and timestamp-based protocols have been widely used to manage locks in relational databases~\cite{bernstein1987concurrency}. However, these approaches are based on static or rule-based policies that do not adapt to dynamic workloads. Recent advances in machine learning have enabled predictive models that anticipate lock requests to mitigate contention and deadlocks. Lock contention is a critical issue in database management, affecting transaction throughput and system efficiency~\cite{tian2018contention, nanda2020comprehensive, wang2021polyjuice}. 

Yu and Jagadish~\cite{yu2014predicting} analyzed resource contention in multitenant databases, highlighting its impact on performance. McWherter et al.~\cite{mcwherter2005improving} examined the online transaction processing (OLTP) locking patterns and introduced the Preempt on Wait (POW) policy to enhance lock management. By statistically characterizing lock wait times, POW selectively preempts lower-priority transactions to reduce contention. This approach enhances lock scheduling efficiency and minimizes delays for high-priority transactions. Tian et al.~\cite{tian2018contention}  presented a novel approach to managing lock contention in transactional databases. The authors introduce the Largest-Dependency-Set-First (LDSF) algorithm, which prioritizes transactions that block the most others, thereby reducing overall transaction latency. The study presents a formal analysis of lock scheduling challenges and demonstrates that implementing LDSF in a real-world database management system can lead to significant performance improvements. Zhou et al. \cite{wang2020predictive} introduced a predictive transaction scheduling approach designed to mitigate lock contention in databases. Their method uses historical workload patterns to forecast transaction conflicts and optimize execution order, thereby reducing the likelihood of contention-related performance degradation.

Gaffney et al. \cite{gaffney2021database} introduced DIBS, a modular transaction isolation service that utilizes optimized predicate locking to replace monolithic storage managers. By locking logical predicates instead of physical data, DIBS reduces overhead and scales to 10.5 million transactions/sec while cutting SQLite writes by 90\% (3x throughput) and MySQL row contention by 40\%. Gaffney et al. demonstrated practical, high-performance isolation via decoupled, predicate-aware design. Wang et al. \cite{wang2021polyjuice} introduced Polyjuice a machine learning approach to optimize database concurrency control by dynamically selecting locking strategies based on transaction clusters and workload patterns. Using reinforcement learning reduces contention by predicting transaction conflicts and adaptively scheduling them, indirectly minimizing unnecessary lock waits.

\subsection{Workload Modeling for Predictive Locking}
Effective workload modeling is essential for predicting database access patterns and optimizing performance. Previous research has explored the characterization of workloads using statistical and machine learning techniques. For example, Martin et al.~\cite{martin2006workload} developed a workload forecasting framework for autonomic database management systems, emphasizing the importance of understanding workload behavior for system optimization.

Building on this foundation, analytical models have been proposed to assess the impact of concurrency control mechanisms on system performance. Di Sanzo et al. \cite{di2010analytical} presented an analytical model of lock-based concurrency control with arbitrary transaction data access patterns, providing insights into how different workloads influence lock contention and overall throughput. Furthermore, Thomasian  \cite{thomasian2024heterogeneous} introduced a heterogeneous data access model for concurrency control, discussing methods to handle high data contention in  OLTP systems. This study highlights the need for realistic workload models to predict lock contention and suggests strategies to mitigate performance degradation under heavy load conditions. Workload management in database systems involves monitoring, managing, and controlling query execution to optimize resource utilization and achieve performance objectives. Zhang et al.~\cite{zhang2017workload} provide a comprehensive taxonomy of workload management techniques in modern database systems, analyzing commercial and research-based approaches while identifying open challenges in workload forecasting and optimization.

Recent advances in self-driving database systems highlight the potential of predictive workload modeling. For example, Ma et al.~\cite{ma2018query} proposed a query-driven workload forecasting approach for self-optimizing database management systems, underscoring the critical role of workload prediction in database performance tuning. Shahrivari et al.~\cite{shahrivari2022workload} presented Adaptive Approximate Query Processing (AAQP), which enhances the query processing by predicting future workloads using Recurrent Neural Networks (RNNs) trained on historical query sequences. It dynamically adjusts data synopses to minimize query execution time while maintaining accuracy. Experiments on real-world workloads show that AAQP effectively optimizes performance, approaching near-optimal execution efficiency.

\subsection{Buffer Pool Optimization and  Prediction}
Efficient buffer pool management is crucial for improving database performance by reducing disk I/O operations. Traditional buffer replacement policies, such as Least Recently Used and Least Frequently Used, have been widely studied~\cite{megiddo2003arc}. However, machine learning approaches have shown promise in outperforming these heuristics.

Li et al.~\cite{li2021drl} introduced DRL-Clusters, a deep reinforcement learning-based approach to buffer pool management in database systems. Their method dynamically adjusts buffer allocations based on workload changes, improving cache efficiency and overall performance. Similarly, Yang et al.~\cite{yang2024rl} proposed RL-CoPref, a reinforcement learning-based coordinated prefetching controller designed to optimize buffer management by predicting future data accesses, ensuring more effective memory utilization.

Furthermore, Zirak et al.~\cite{zirak2023selep} developed SeLeP, a learning-based semantic prefetcher tailored for exploratory database workloads. By analyzing data access patterns, SeLeP enhances prefetching accuracy, leading to improved buffer pool utilization and reduced query latency. These studies demonstrate the effectiveness of predictive models in optimizing memory utilization, aligning with the research goals in forecasting table and page locks to enhance concurrency control in database systems.

Unlike this study, the mentioned studies focus on lock management strategies rather than explicit lock sequence prediction. In the context of IBM Db2, prior studies have examined statistical models and heuristics for lock contention analysis~\cite{zhang2012discovering}. However, deep learning-based lock prediction remains an underexplored area, particularly for predicting table and page level locks.
\section{Methodology} \label{sec:methodology}

Our methodology is designed to develop and evaluate deep learning based models for predicting database locks in IBM Db2. This section outlines the key steps involved, including data collection, preprocessing, model architecture, training, and evaluation.

\begin{table}[htbp]
\centering
\caption{Distribution of Lock Object Types Collected in Db2}
\begin{tabular}{@{}lrr@{}}
\toprule
\textbf{Lock Type} & \textbf{Record Count} & \textbf{Percentage (\%)} \\
\midrule
\textbf{PAGE}          & 2,702,900     & 31.00                  \\
CATALOG       & 2,213,913     & 25.40                  \\
\textbf{TABLE}         & 2,178,913     & 24.99                  \\
VARIATION     & 669,924       & 7.68                   \\
PLAN          & 642,945       & 7.38                   \\
SEQUENCE      & 300,970       & 3.45                   \\
INTERNAL      & 8,254         & 0.09                   \\
TABLESPACE    & 4             & 0.00                   \\
\midrule
Total         & 8,717,823     & 100.00                 \\
\bottomrule
\end{tabular}
	\label{tab:LockseDataset1}
\end{table}

\subsection{Data Collection}
We collect database lock traces from an IBM Db2 environment running HammerDB~\cite{hammerDBLink} TPC-C benchmark workload v.~5.11~\cite{tpcc_v511} while enabling the Db2 trace using the "db2trc" command. TPC-C is a widely used OLTP benchmark designed to simulate a high-volume transaction system, such as an e-commerce or inventory management system. The collected lock traces are for eight tables: customer, district, history, neworder, orderline, orders, stock, and warehouse. The orderline table is the one with the highest number of associated lock data.  The trace explicitly captures Db2 functions related to lock operations, including "sqlplrq" (lock request), "sqlplrl" (lock release), and "sqlplrem" (lock removal). For the experimental workload, we apply a TPC-C benchmark configuration with 100 warehouses, 10 asynchronous clients, and 100 virtual users, generating realistic transactional activity. We collect around 25GB of a trace log that is ready for further processing.

\subsection{Data Preprocessing}
The collected raw lock trace log undergoes several pre-processing steps to extract different information about each log lock. Pre-processing is performed by parsing the trace log and extracting each required element into a structured format for subsequent steps. Table~\ref{tab:LockseDataset1} presents the distribution of the lock types extracted from the processed trace log. Below, we highlight a list of the extracted data for each lock: 

\begin{itemize}[leftmargin=*]
 \item   \textbf{Lock ID: } A unique identifier assigned to each lock instance in Db2. This helps track individual lock operations and distinguish between different locks in the system.
\item  \textbf{Start Time: } The timestamp that indicates when the lock was acquired.
\item  \textbf{End Time: } The timestamp that indicates when the lock was released.
\item  \textbf{Mode: } Specifies the type of lock acquired, such as Shared (S), Exclusive (X), Intent Exclusive (IX), or Update (U)~\cite{IBMDB2Modes}. The mode determines the level of access allowed while preventing conflicts with other transactions.
\item  \textbf{Lock Object: } The database resource that is being locked, which can be a data page, table, tablespace, or an index. Identifying the locked object helps to understand the contention of the transactions and access patterns.
\item  \textbf{Page ID: } This indicates the specific database page that is locked. Pages typically contain multiple rows. Page-level locking strikes a balance between concurrency and overhead.
\item  \textbf{Table Name: } The name of the table associated with the lock. Knowing which tables are affected by locking can help optimize queries, indexing strategies, and database design to minimize contention. The table-level locks include a table name, while the page-level locks include a table name and a page ID. 
\end{itemize}

In this study, we focus on the table and page locks, which cover 50\% of the lock types from Table~\ref{tab:LockseDataset1}. To prepare the data for model training, we categorize the lock data into two main groups: (1) Table Locks and (2) Page Locks. Table locks comprise all Db2 lock records where the "Lock Object" is a table, while the page locks encompass those where the "Lock Object" is a data page.  To ensure our analysis focuses on the TPC-C workload, we filter out system locks associated with the "SYSIBM" schema. After applying this filter, we obtain 1.54 million table locks and 2.66 million page locks.

For each lock type (i.e., Table and Page locks), we construct sequential transaction timelines by organizing lock events into ordered sequences. We sort the locks data by the start time in ascending order. Feature encoding is applied to transform page IDs and table names into numerical representations. Continuous attributes, such as timestamps, undergo normalization, and a time-based train-test split is used to evaluate the model on future transactions.

\subsection{Model Architecture}
\begin{figure}[t]
    \vspace{-6mm}
    \includegraphics[trim=2 7 2 2, clip, width=0.99\linewidth]{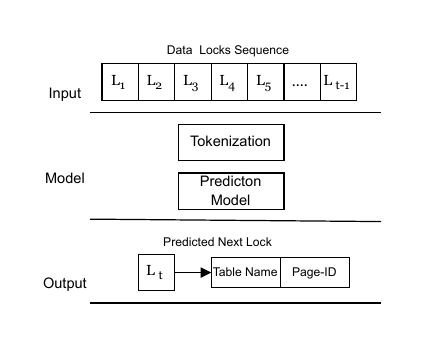}
           \vspace{-4mm}
    \caption{Predicting the next lock in a sequence.}
    \label{fig:Prediction}

\end{figure}

In this study, we use deep learning architectures, such as Transformer~\cite{DBLP:conf/nips/VaswaniSPUJGKP17} and LSTM~\cite{DBLP:journals/neco/HochreiterS97}, to predict the next locks in a Db2 database system. Figure~\ref{fig:Prediction} presents an overview of the prediction of the next lock based on an input of the lock sequence.
For both models, the input sequence consists of 25 locks each comprised of one or two tokens (Table $T$ and Page ID, respectively), which are first transformed into 128-dimensional dense vector representations using an embedding layer. In the Transformer model, positional encoding is added to these embeddings to preserve the order of tokens in a sequence. The core of the architecture features a Transformer encoder layer that utilizes multi-head self-attention to capture complex, long-range dependencies among lock events. The encoder-only Transformer model we use consists of 8 self-attention heads with a hidden dimension of 512 and a dropout rate of 10\%. In contrast, the LSTM model processes the embedded sequences using a recurrent layer~\cite{sherstinsky2020fundamentals} capable of learning temporal patterns and dependencies through gated memory cells, which help retain context over longer sequences. Their recurrent nature also makes LSTM models particularly effective for sequential prediction tasks involving variable-length dependencies. The LSTM model uses a hidden dimension of 256 in each gated memory cell.

In both architectures, encoded lock representations are passed through a feedforward network, including a dense layer with Rectified Linear Unit (ReLU)~\cite{nassiri2023transformer, sherstinsky2020fundamentals} activation, and culminate in a softmax output layer that predicts the next page ID and table name to be locked. This dual-model setup allows us to compare the effectiveness of attention-based and recurrent-based architectures in the context of database lock prediction.

\subsection{Model Training}
\label{subsection:ModelTrain}
 
The models were implemented using Python v.~3.11.5, Keras v.~3.3.3, and the TensorFlow v.~2.16.1 backend. To train the model, we preprocess the lock sequences by tokenizing and padding them to a fixed length. A single tokenizer is used for both input and output sequences, ensuring consistent vocabulary mapping. The dataset is chronologically divided into a test set, comprising 30\% of the data, and a train set. A validation set, consisting of 20\% of these initial training data, is then partitioned off, with the remainder used for model training. This chronological partitioning preserves the temporal order of lock requests to prevent data leakage.

All models were trained using categorical cross-entropy loss and the AdamW optimizer~\cite{DBLP:conf/iclr/LoshchilovH19}, with a constant learning rate of $0.001$ and the remaining parameters set to the Tensorflow defaults. The training process involves iterating over the dataset in mini-batches of size 32. To enhance generalization, we incorporate dropout regularization to prevent overfitting. Additionally, a model checkpoint was used to monitor the validation loss for each epoch. The model weights that achieved the lowest validation loss in all 30 epochs were saved and used for the final evaluation. To ensure the robustness of our findings, each experiment was executed for a total of 10 iterations.

For the lock prediction, we adopt two modeling approaches: Global Models and Local Models. We define these model types as follows.

\begin{itemize}[leftmargin=*]
\item \textbf{Global Model:} A single unified model is trained using data from all tables and predicts the next lock, regardless of the table. For example, in the case of page-level lock, this model learns to predict the next (Table, Page ID) combination in a sequence, enabling it to generalize across different tables and capture inter-table patterns.

\item \textbf{Local Model:} For this model type, a separate model is trained for each individual database table and tested on the same table. For example, for Table $T$, we train a model using all data from Table $T$ and predict the next locks for Table $T$. This is an appropriate type for the page-level locks, where a model is responsible for predicting the next Page ID within its respective table. This approach allows the model to specialize in the unique access patterns and characteristics of each table.
\end{itemize}

\subsection{Model Evaluation}

Performance evaluation is performed using multiple metrics, including accuracy, precision, recall, and F-measure, to assess how effectively a model anticipates locking sequences. In the following, we describe the performance metrics~\cite{diederik2014adam}:

\begin{itemize}[leftmargin=*]
\item \textbf{Accuracy:} Measures the correctness of predicted locks.

\item \textbf{Precision:} Measures how many of the predicted locks for a specific table or page ID are actually correct. It is defined as \( \text{Precision} = \frac{\text{TP}}{\text{TP} + \text{FP}} \), where TP and FP denote the number of true positives and false positives, respectively.

\item \textbf{Recall:} (also known as sensitivity) Evaluates how many of the actual lock events for a specific table or page ID were correctly predicted. It is calculated as \( \text{Recall} = \frac{\text{TP}}{\text{TP} + \text{FN}} \), where FN represents the number of false negatives.

\item \textbf{F1-Score:} The F1-Score is the harmonic mean of precision and recall, defined as \( \text{F1} = 2 \times \frac{\text{Precision} \times \text{Recall}}{\text{Precision} + \text{Recall}} \).

\end{itemize}

These metrics are especially important in multi-class classification tasks like this study, where the number of classes (tables or page IDs) is large. High precision ensures the model does not incorrectly predict locks for the wrong table or page ID (minimizing false alarms), while high recall ensures that the model captures most of the actual lock events (minimizing missed locks).  A high F1-Score indicates that the model achieves a good balance between correctly predicting lock events (recall) and minimizing incorrect predictions (precision).

To assess a model's ability to forecast locking sequences across multiple future steps, we use the term \emph{Horizon}. In this study, we evaluate up to four horizons. For instance, a horizon of 3 indicates the model’s accuracy in predicting the next three consecutive locks. By default, predicting only the immediate next lock corresponds to Horizon~=~1. Overall, the predictions are compared against expected outputs, and the softmax probabilities are analyzed to interpret the model's confidence. 

To ensure robustness and account for variance due to random initialization, we train each deep learning model (Transformer and LSTM) over 10 iterations, each using a different random seed. For each prediction horizon (e.g., Horizon 1 through Horizon 4), performance metrics such as accuracy, precision, recall, and F1-score are computed in each iteration. The final reported value for each metric at a given horizon is obtained by averaging the results over the 10 runs. This approach helps smooth out anomalies due to random factors and provides a more reliable estimate of model performance per horizon.

\section{Results}

\subsection*{RQ1: \textbf{\RQOne}}

\noindent \textbf{Motivation.}   Achieving holistic database encryption without incurring downtime or excessive storage overhead remains a critical challenge for companies striving to comply with regulations such as GDPR, HIPAA, and PCI DSS~\cite{rjaibi2018holistic, folorunso2024security} as well as to ensure quantum-safety. A major bottleneck in this process is the need to encrypt or re-encrypt existing databases without disrupting active transactions. Predicting which table will be locked next is essential to integrate encryption seamlessly into live database operations.
Accurate predictions allow for proactive encryption strategies that minimize performance degradation, such as aligning encryption operations with predicted lock sequences to avoid redundant re-encryption or unnecessary delays~\cite{ibm_patent}. Furthermore, forecasting table-level locks is key to ensuring that online encryption processes do not interfere with critical workloads. In large-scale database systems such as IBM Db2, this predictive capability represents a meaningful step toward enabling real-time, online encryption without downtime or significant storage overhead.

\begin{table}[t]
\centering
\caption{Summary for the performance metrics averages for all the  Table Lock 
predictions across the Studied Models for Horizon~=~1. The reported values represent the average of the ten modelling repetitions across the metrics.}
\centering
\begin{tabular}[t]{@{}lllll@{}}
\toprule
\textbf{Model} & \textbf{Accuracy}& \textbf{Precision} & \textbf{Recall} & \textbf{F1} \\
\midrule
Transformer & 0.475 & 0.488 & 0.359 & 0.455\\
LSTM & 0.488 & 0.496 & 0.373 & 0.470\\
Naive & 0.084 & 0.039 & 0.039 & 0.048 \\
\bottomrule
\end{tabular}
   \label{tab:RQ1TableLocks}
\end{table}

\noindent \textbf{Approach.} To address the challenge of table-level lock prediction, we structured historical lock sequences to capture dependencies between table accesses. We trained various models on all tables' lock data to learn cross-table dependencies and generalize across the entire database. The trained models leverage the Transformer and LSTM architectures described in Section~\ref{sec:methodology}, utilizing self-attention mechanisms to model long-range dependencies in lock sequences.

To evaluate performance, we compared the trained models against a native baseline that assumes the next lock is identical to the previous lock: a simple yet competitive heuristic in lock prediction tasks.  Additionally, we extended our analysis beyond single-step prediction to study the model’s performance across varying prediction horizons, such as predicting the next 2, 3, or 4 locks. This multi-horizon evaluation offers insights into the model’s ability to forecast longer sequences, which may be crucial for optimized online database encryption.

\begin{figure}[t]
    \includegraphics[width=0.48\textwidth]{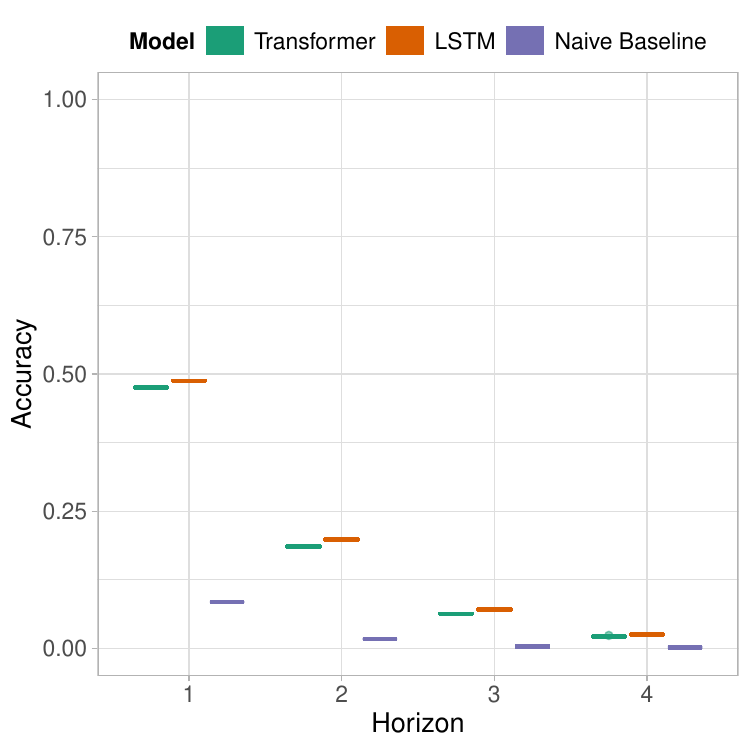}
    \caption{This figure presents the table-level prediction accuracy across different tables and horizons with a comparison of local models and naive models. The model here predicts the sequence of the next Table $T$ to be locked.  The figure shows that the deep learning approach outperforms the Naive Baseline.}
     \vspace{-4mm}
    \label{fig:RQ1nativegloballocal}
\end{figure}

\noindent \textbf{Results.} We discuss the results of this RQ as presented prediction results in Table~\ref{tab:RQ1TableLocks} and Figure~\ref{fig:RQ1nativegloballocal}.

\begin{figure*}[t]
    \vspace*{-2mm}
    \includegraphics[width=1.0\textwidth]{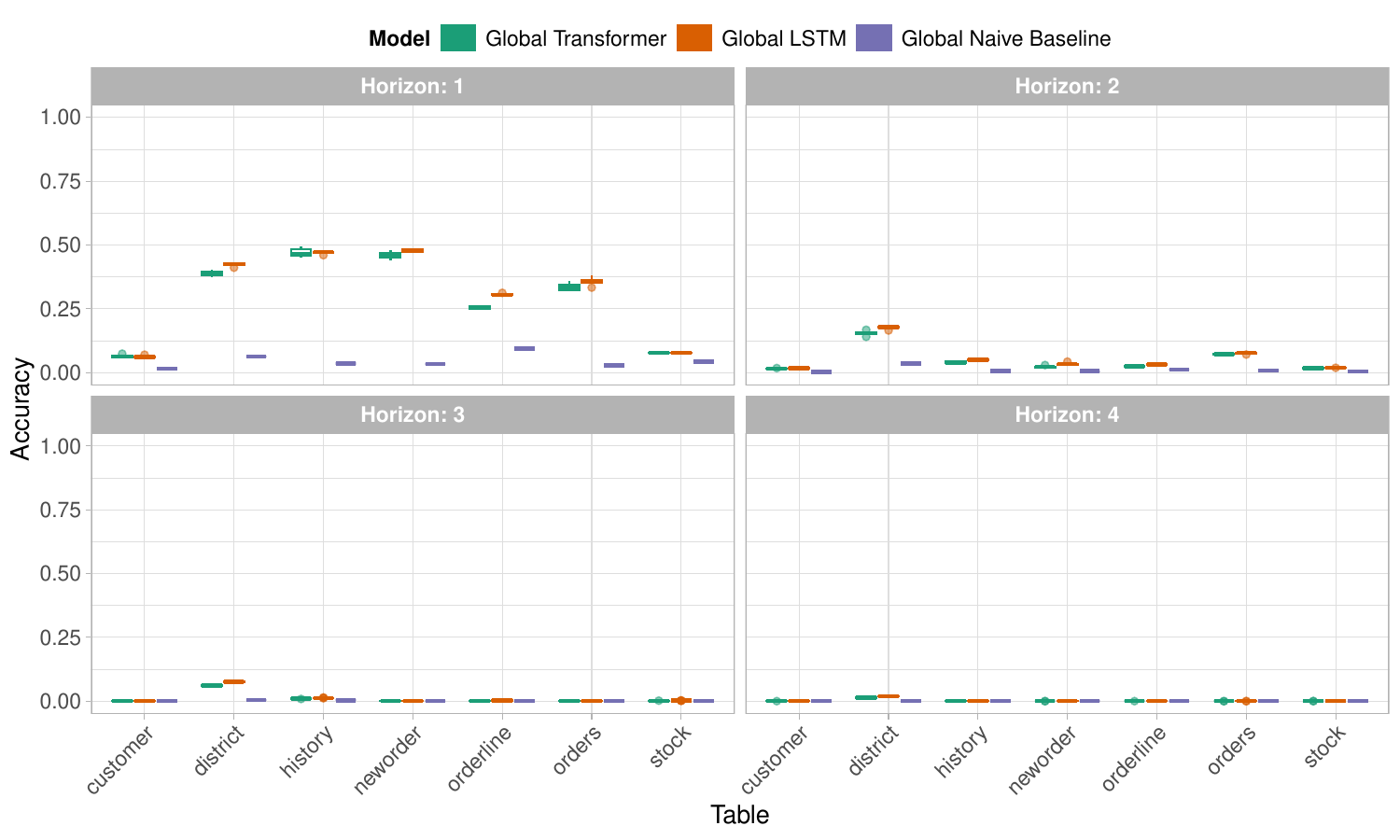}
    \caption{This figure presents the page-level prediction accuracy across different tables and horizons. The figure compares the performance of the global models across all tables and horizons.  The Transformer and LSTM models outperform the Naive Baseline across all the tables and horizons. The model here predicts a sequence of next locks as a combination (Table $T$, Page ID Bin).  }
    \label{fig:RQ2globallocal}
\end{figure*}

\begin{figure*}[t]
    \vspace*{-2mm}
    \includegraphics[width=1.0\textwidth]{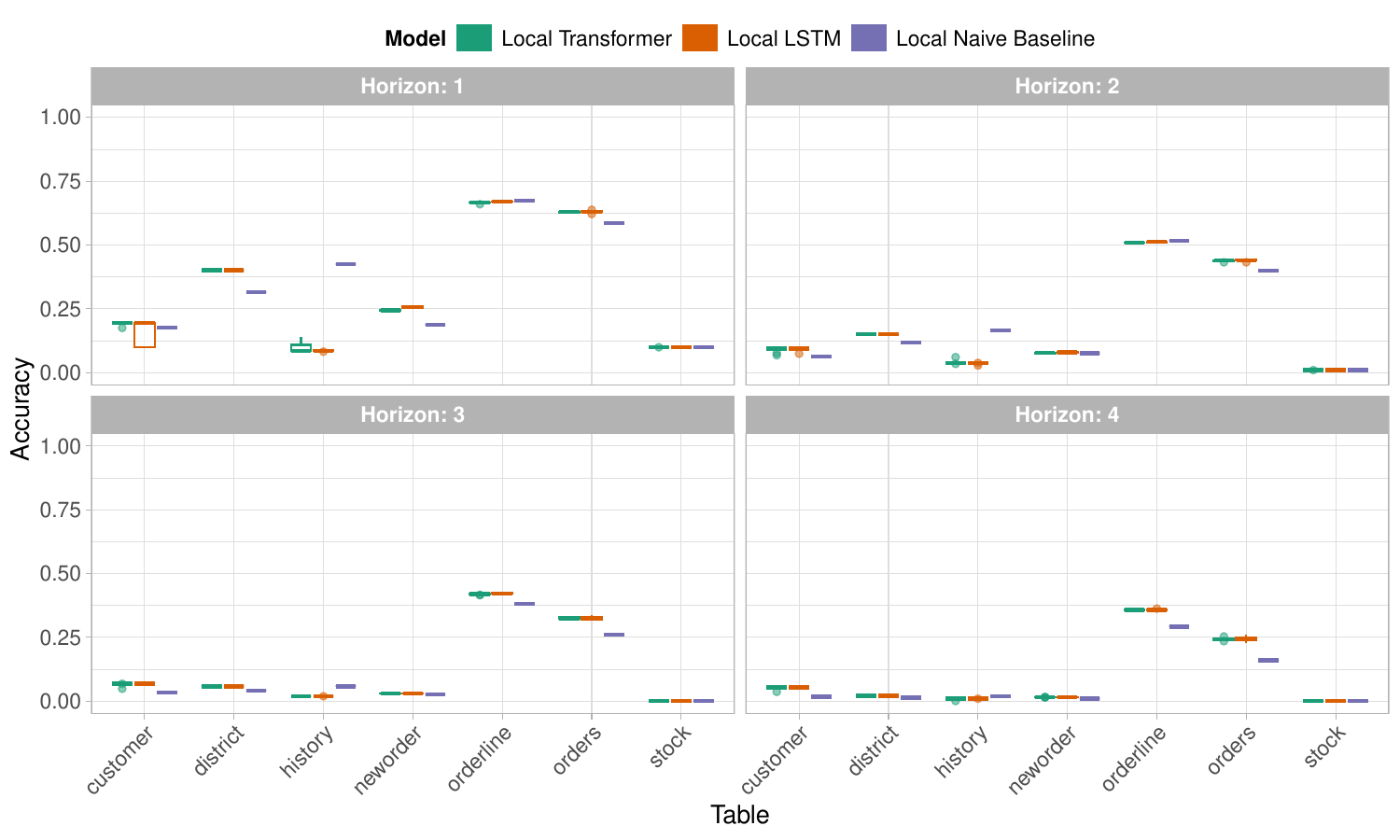}
    \caption{This figure presents the local models' prediction accuracy for page-level across different tables and horizons. The figure compares the prediction for the Transformer and LSTM models versus the Naive Baseline. The results show that the Naive Baseline outperforms the deep learning models in some cases. The model here predicts a sequence of next locks as Page ID Bin. Predicting only the next lock Page ID Bin is less challenging than the Global Model (compare with the Global models' Figure~\ref{fig:RQ2globallocal}).}
    \label{fig:RQ2RowIDPrediction}
\end{figure*}

\noindent \textbf{\textit{The prediction models consistently outperform the Naive Baseline.}} 
Table~\ref{tab:RQ1TableLocks} presents the performance metrics for both deep learning models (Transformer and LSTM) compared to a Naive Baseline predictor.   The results clearly show that both the Transformer and the LSTM models significantly outperform the Naive Baseline across all metrics. Deep learning models achieve higher accuracy, indicating a better overall prediction of the next lock. They also demonstrate improved precision and recall, suggesting that they are more effective at both: correctly identifying relevant locks and minimizing false positives or missed predictions. The F1-Measure, which balances precision and recall, further confirms the superior performance of the deep learning models versus the Naive Baseline.

These improvements validate the hypothesis that deep learning-based models can effectively leverage the sequential nature of database locks to make more informed predictions. In particular, the consistent superiority across multiple evaluation metrics underscores the utility of incorporating temporal patterns (captured by the Transformer’s self-attention mechanism and LSTM’s memory units) into predictive models for database lock behaviour. The results indicate that both the Transformer and LSTM models have the potential to serve as practical components of intelligent DBMS subsystems, particularly for online encryption of enterprise databases like DB2 without downtime or storage overhead.

\noindent \textbf{\textit{The prediction performance is consistent across all horizons.}}
Figure~\ref{fig:RQ1nativegloballocal} provides a comparative analysis of predication performance across various horizons. The results clearly indicate that both the Transformer and LSTM models significantly outperform the Naive Baseline across all prediction horizons (e.g., Horizon 1, Horizon 2) and all evaluated tables. The evaluated deep learning models demonstrate a strong and consistent ability to anticipate future locks, even as the temporal distance between the observed and predicted locks increases.

For example, at Horizon 1, where the goal is to predict the immediate next lock, the Transformer model achieves an average accuracy exceeding that of the Naive Baseline, particularly on critical tables such as \dbtable{customer}, \dbtable{history}, and \dbtable{orders}. In contrast, the Naive Baseline struggles to reach 9.1\% accuracy in most cases. This performance gap becomes even more pronounced at Horizons 2-4, where the models' accuracy deteriorates. However, the Transformer and LSTM models retain a higher level of predictive accuracy, illustrating their capacity to learn and model temporal dependencies beyond the immediate future. Overall, the results reinforce the value of using sequence-aware deep learning models in lock prediction tasks. Unlike the Naive Baseline, which lacks memory or pattern recognition.

\subsection*{RQ2: \textbf{\RQTwo}}

\noindent \textbf{Motivation.} While predicting table-level locks helps to optimize database-wide encryption strategies, page-level locks (represented by page IDs in Db2) offer finer control over encryption timing and transaction efficiency. In enterprise environments, where minimizing downtime and storage overhead is crucial for enabling encryption on existing databases, accurately forecasting the next page to be locked can help schedule encryption operations without interfering with active transactions. Page-level encryption must be carefully aligned with ongoing access patterns to avoid excessive performance overhead, redundant encryption operations, and unnecessary lock waiting times. By predicting the next data page to be locked, the database can preemptively encrypt data pages just before they are accessed, ensuring security policies are enforced without impacting system responsiveness. This RQ research is a crucial step toward achieving
fine-grained online encryption without downtime or space overheads.

\noindent \textbf{Approach.} 
Building on the methodology of RQ1, we extend our deep learning–based framework to predict page-level locks represented by page identifiers (Page IDs) in Db2. Unlike RQ1, which focuses on table-level dependencies, this approach structures lock sequences to include both the table name and data page as tokens, enabling the model to learn fine-grained locking patterns. Specifically, the input and output sequence consists of pairs (Table $T$, Page ID Bin), and the model predicts the next token in this sequence (see Figure~\ref{fig:Prediction}). We developed two distinct models:  (1) a global model trained on all tables to learn cross-table dependencies and generalize throughout the database, and (2) a local model trained individually for each table to capture table-specific locking patterns (see Section~\ref{subsection:ModelTrain}). Both models leverage Transformer architectures, utilizing self-attention mechanisms to model long-range dependencies in lock sequences. Out of the eight tables mentioned in Section~\ref{sec:methodology}, we filter out the Warehouse table since it has only a single page ID value for all the page-level locks. 

To enhance prediction performance and reduce the complexity of the output space, we applied a binning strategy to the page IDs. Instead of predicting raw page ID values, we grouped them into 10 bins based on their distribution within each table. This transformation reduces noise and enables the model to generalize better across similar page access patterns, particularly in cases where the page ID values are sparse or highly variable. This binning-based formulation enables the models to capture page-level locking behaviour more effectively, while keeping the prediction space tractable and improving model convergence in diverse database workloads. For performance evaluation, we use the same metrics as in RQ1. Additionally, we conduct a multi-horizon analysis to assess the model’s ability to predict sequences of 2, 3, or 4 locks.

\noindent \textbf{Results.} The Results of this RQ are illustrated in Figures~\ref{fig:RQ2globallocal} and \ref{fig:RQ2RowIDPrediction}. We discuss the findings below.

\begin{figure*}[t]
    \vspace*{-2mm}
    \includegraphics[width=1\textwidth]{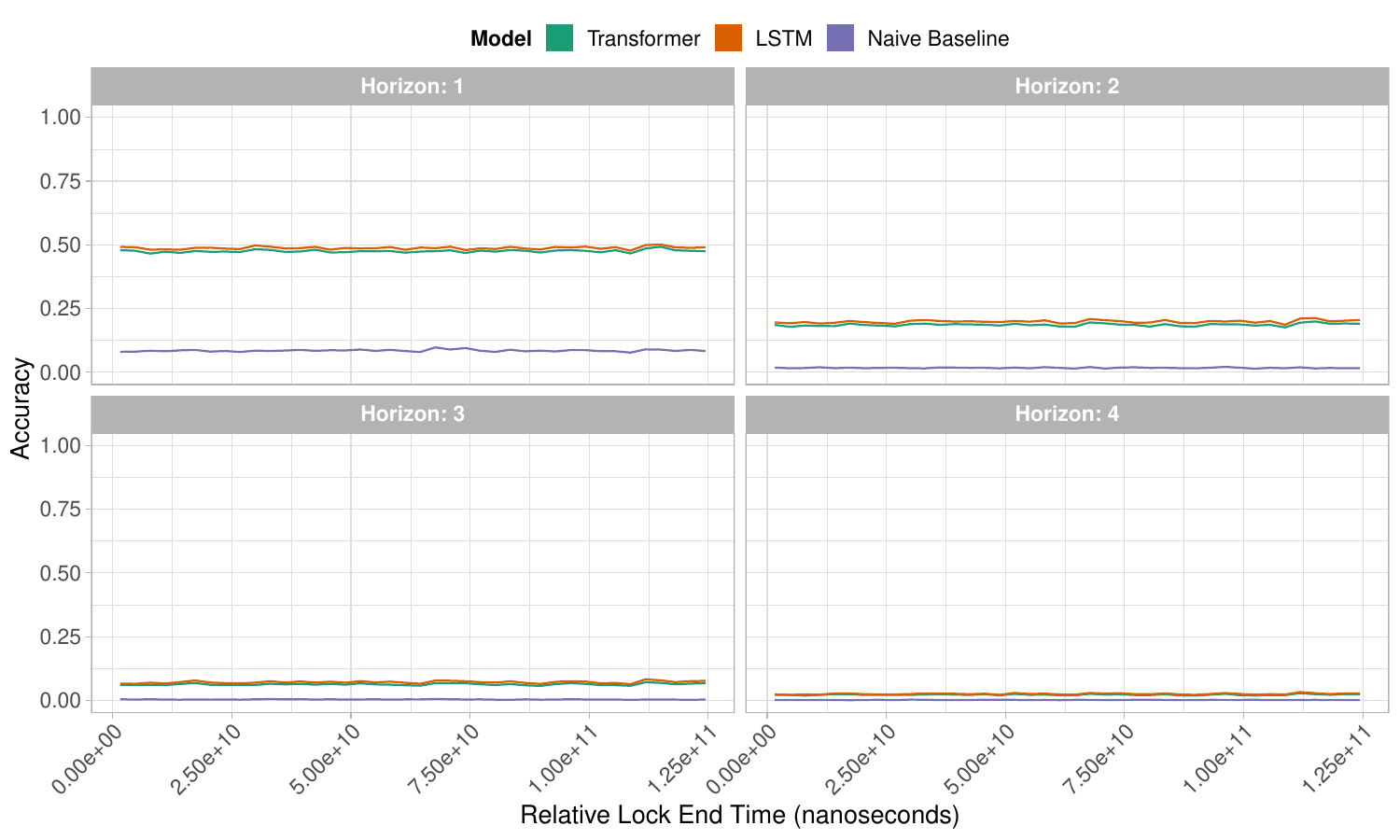}
    \caption{This figure presents the  \textbf{table-level} prediction accuracy over the workload period and horizons. The figure highlights whether the performance accuracy changes over time. As a table-level prediction model, the model here predicts the sequence of locks as the next Table $T$ to be locked.}
    \label{fig:RQ3Overtime1}
\end{figure*}

\begin{figure*}[t]
    \vspace*{-2mm}
    \includegraphics[width=1\textwidth]{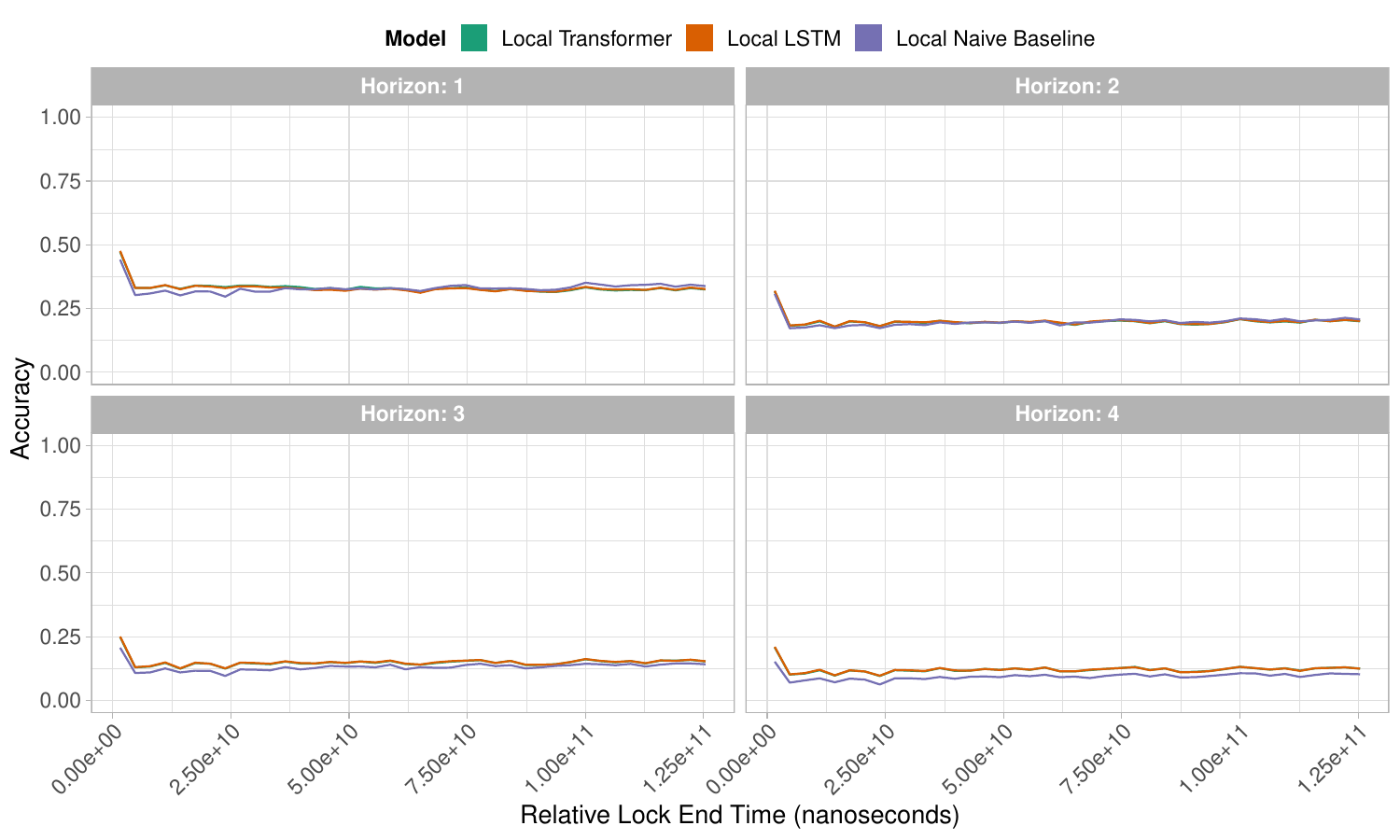}
    \caption{This figure presents the local \textbf{page-level} prediction accuracy over the workload period and horizons. The figure highlights whether the performance accuracy changes over time.}
    \label{fig:RQ3Overtime1Row}
\end{figure*}

\noindent \textbf{\textit{The Global deep learning models outperform the Naive Baseline. }} 
Figure~\ref{fig:RQ2globallocal} presents the page-level prediction accuracy across different tables and prediction horizons. The results clearly show that both the Global Transformer and Global LSTM models consistently outperform the Naive Baseline across all tables and horizons, particularly at Horizon 1. At this short horizon, both deep models achieve noticeably higher accuracy for most tables, indicating their strong ability to model immediate next-lock patterns.

As the prediction horizon increases (Horizons 2–4), accuracy declines for all models, reflecting the growing difficulty of longer-term sequence prediction. However, even as performance drops, the deep learning models, especially the LSTM model, continue to outperform the Naive Baseline in almost all cases. The gap between the deep models and the Naive Baseline is especially prominent for tables such as \dbtable{district}  and \dbtable{orders}, where the deep learning models demonstrate higher accuracy and robustness.

Notably, the variance of the deep models remains relatively low at Horizon 1 but increases slightly with increasing horizon depth, although it remains generally more stable than the Naive Baseline. No table shows the Global Naive Baseline outperforming the deep models at any horizon, suggesting that learning temporal and contextual locking patterns provides a strong advantage. These findings reinforce the suitability of deep learning models for predicting next-lock events at a global scope, with the Transformer and LSTM showing particular promise due to their consistent performance across both tables and horizons. One explanation for the low performance at large horizons (2-4), is due to the fact that the models predict a pair (Table $T$, Page ID) and not a single value as in RQ1.

\noindent \textbf{\textit{Surprisingly, for a Local Model, the Naive Baseline shows a competitive performance.}}
Figure~\ref{fig:RQ2RowIDPrediction} presents the page-level prediction accuracy across all tables and four using local models. In this case, every table will have its own trained model, and the prediction would focus only on the Page ID Bin (because the table name $T$ in the ``Table $T$, Page ID Bin'' tuple stays constant).  While the Local Transformer and Local LSTM models generally show improved accuracy over the Naive Baseline in structured tables (such as \dbtable{orderline} and \dbtable{orders})~---~especially at Horizon 1~---~the Naive Baseline often remains surprisingly competitive, and in some cases, even outperforms the deep models. For example, at Horizon 1, the Naive Baseline performs similarly or better than both deep models on \dbtable{district}, \dbtable{history}, and \dbtable{neworder}. This trend persists and becomes more apparent in longer horizons (Horizon 2 to 4), particularly in low-variability or highly repetitive tables.

As the prediction horizon increases, all models exhibit a decline in accuracy due to the increased complexity of forecasting multiple future page accesses. However, the performance of deep models does not always scale favorably with horizon length. In some tables (e.g., \dbtable{district}, \dbtable{history}, \dbtable{stock}), the Local Naive Baseline maintains or exceeds the performance of both the Transformer and LSTM models across all horizons. This highlights that, in the local context, simple recency-based heuristics can be robust and effective, especially under consistent or repetitive locking behavior.

Overall, these results emphasize the unexpected resilience of the Naive Baseline in local sequence prediction for Page ID Bins. Despite its simplicity and lack of learning overhead, it can rival and sometimes outperform more complex models, particularly in scenarios characterized by stability and routine. This suggests that for certain OLTP environments, lightweight heuristics may offer a practical and resource-efficient alternative to deep learning approaches.

\subsection*{RQ3: \textbf{\RQThree}}

\noindent \textbf{Motivation.}  Locking patterns in a database system are not static; they evolve based on workload fluctuations, query patterns, and system optimizations. Understanding how lock predictions shift over time is crucial for ensuring the long-term reliability of predictive models, particularly in the context of enabling live encryption with zero downtime and storage overhead. If lock sequences exhibit significant temporal drift, models may require adaptive learning mechanisms to maintain accuracy and effectiveness. Additionally, analyzing temporal changes in lock predictions can help identify workload trends, detect anomalies, and refine encryption scheduling strategies to minimize performance impact. By addressing this research question, we aim to evaluate the robustness of lock prediction models over time and explore a possible need for continuous optimization of the trained lock prediction models.

\noindent \textbf{Approach.} To analyze how lock predictions evolve over time, we explore the separate models for table-level (see RQ1) and page-level locks (see RQ2), capturing their distinct access patterns. This RQ methodology involves segmenting the DB2 lock traces into time-based windows, allowing us to assess temporal shifts in prediction accuracy and model performance. We track key metrics discussed in Section~\ref{sec:methodology},  across different time periods to identify potential drift in locking patterns. To further investigate variations, we compare the models' performances across different horizons.

\begin{figure}[t]
    \vspace*{-2mm}
    \includegraphics[width=0.48\textwidth]{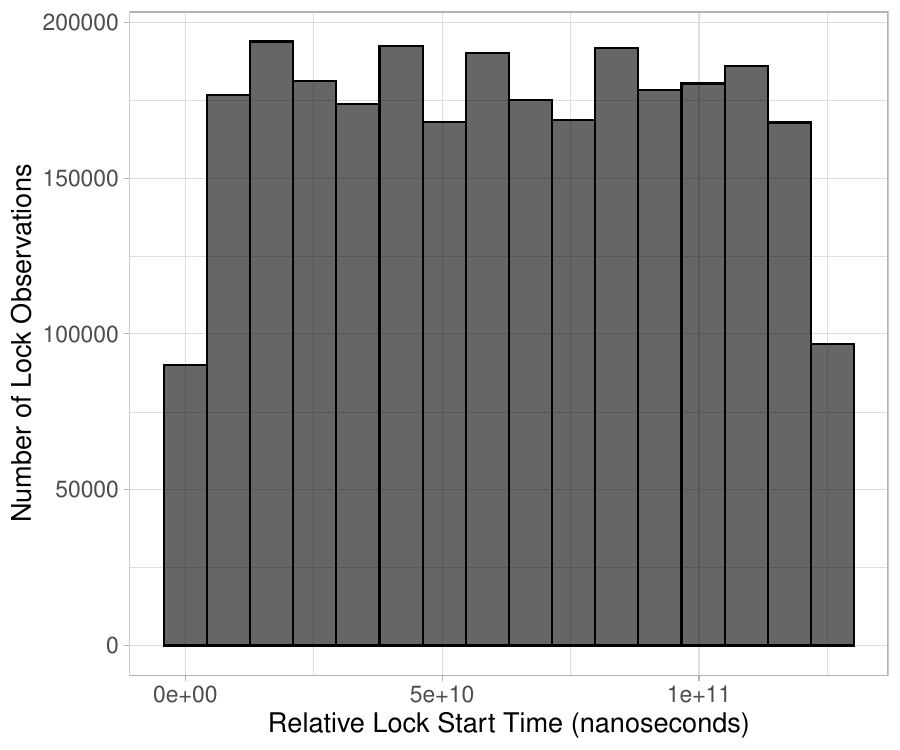}
    \caption{Distribution of table-level testing observations over time (in nanoseconds) based on relative lock start time. The histogram shows the number of recorded table lock instances in fixed-width time intervals for horizon = 1. The relatively uniform distribution indicates consistent lock activity across the monitored testing period, with slight variations near the boundaries.}
    \label{fig:RQ3ObservationDist}
\end{figure}

\noindent \textbf{Results.}  The Results of this RQ are illustrated in Figures~\ref{fig:RQ3Overtime1} and~\ref{fig:RQ3Overtime1Row}. We discuss the findings below.

\noindent \textbf{\textit{Table level prediction exhibits a temporal stability.}} Figure~\ref{fig:RQ3Overtime1} illustrates the table-level prediction accuracy over time and across the four forecasting horizons (Horizon 1 to 4). The x-axis represents the relative lock end time in nanoseconds, while the y-axis denotes the percentage of correctly predicted locks within each time step. Each subplot corresponds to a specific prediction horizon. Across all horizons, the temporal prediction performance remains remarkably stable over time for all three models: Transformer, LSTM, and the Naive Baseline. No substantial drops or spikes in accuracy are observed, even as the workload progresses, suggesting that locking patterns do not experience strong concept drift over time.  Some fluctuations occur only near the end of the timeline, where the number of data points diminishes. Figure~\ref{fig:RQ3ObservationDist} presents a data distribution example over time, highlighting the fluctuations at the start and end of the testing period.  These results suggest that the models, once trained, generalize effectively over time and do not require frequent retraining to sustain performance.

Importantly, the relative ranking among models is also consistent throughout time and across horizons. LSTM and Transformer models significantly outperform the Naive Baseline, especially at Horizons 1 and 2, maintaining a ~50\% accuracy rate in Horizon 1 versus ~10\% for the Naive Baseline. Even at longer horizons (3 and~4), the gap remains evident, albeit at lower absolute accuracy levels, indicating increased prediction difficulty.

Overall, the prediction results demonstrate both accuracy and temporal resilience, reinforcing their potential practicality for online encryption under dynamic workloads.

\noindent \textbf{\textit{Page-level prediction maintains temporal consistency.}}
Figure~\ref{fig:RQ3Overtime1Row}  presents the page-level prediction accuracy across the workload period for Horizons 1 and 4. Similar to  Figure~\ref{fig:RQ3Overtime1}, the x-axis represents relative lock end time in nanoseconds, while the y-axis shows prediction accuracy. Three models are compared: Local LSTM, Local Naive Baseline, and Local Transformer. The results demonstrate stable temporal performance for all models, with no significant fluctuations in accuracy as the workload progresses. We notice that the deep learning models maintain a performance advantage throughout the entire observation period, consistently outperforming the Naive Baseline model. The big drop in performance that happens at the beginning of the timeline  is due to a shortage of data points at the start of the workload (See Figure~\ref{fig:RQ3ObservationDist} as an example of workload distribution). Due to space constraints, we omit the figure for global page-level models; however, they exhibit similar trends, further supporting the observation of temporal stability in lock prediction. This consistency suggests that page-level locking patterns remain relatively stationary over time, showing no clear signs of concept drift. As for table-level prediction, these results highlight that page-level lock prediction shares the temporal stability characteristics while maintaining the consistent performance advantage of deep learning models across all horizons.

\section{Implications}
The results of our study demonstrate the feasibility of using deep learning models for database lock prediction, offering several practical implications for database performance optimization, security, and resource management.

\begin{enumerate}[leftmargin=*]
\item Improved Concurrency Control: By accurately predicting which data pages and tables will be locked next, database management systems (DBMS) can implement proactive concurrency control strategies. This can reduce contention, minimize transaction delays, and lower the likelihood of deadlocks. Future work could explore adaptive locking mechanisms that adjust transaction scheduling based on predicted lock conflicts.
\item Optimized Buffer Pool Management: The study results suggest that predictive caching can be used to preload frequently accessed or soon-to-be-locked data pages into memory, reducing disk I/O and improving query responsiveness. Similarly, predictions can prevent premature eviction of pages that are likely to be reused shortly. These capabilities offer strong implications for buffer pool optimization in systems like Db2, aiding with data-driven prefetching and eviction policies based on lock access patterns. However, it is important to note that our work focuses primarily on write operations, whereas buffer pool management typically considers both reads and writes. As such, more research is needed to fully integrate predictive locking into comprehensive buffer pool strategies.  
\item Enhanced Security and Access Control: In security-sensitive environments, predicting locks can help optimize online encryption and decryption of sensitive database content. By anticipating access patterns, encryption overhead can be reduced, ensuring that critical data remains protected without excessive performance costs. Future research could integrate lock prediction with dynamic security policies, adjusting encryption strategies based on predicted access patterns.
\item Generalizability to Other DBMS and Workloads: Although our study focuses on IBM Db2, the methodology may be extended to other RDBMS, such as MySQL, MS SQL, Oracle DB, and PostgreSQL. Potentially, the approach can be adapted to different database workloads, including OLTP, OLAP (Online Analytical Processing), and hybrid workloads.
\item Potential Integration with AI-Driven Optimization: With the rise of AI-assisted database management, lock prediction models could be integrated into self-tuning database systems. By combining lock prediction with other AI-driven optimizations, such as query plan adjustments and index tuning, databases can achieve higher autonomy and efficiency~\cite{ma2018query,kim2021deepdb}.

\item Trade-off between using local and global models: A global model simplifies the machine learning operations by requiring only one model to be trained and maintained, which can then be applied across all tables. This approach reduces complexity and overhead in deployment and monitoring. However, local models typically offer better performance, as they are tailored to the specific characteristics of each individual table. The downside is that each table requires its own separate model, increasing the workload for training, deployment, and ongoing maintenance.

\item Naive vs. Deep Learning Model: As shown in the RQ3 answer, the Naive Baseline sometimes performs only slightly worse than more sophisticated machine learning models. This makes it an attractive option for integration into the database engine code, as it introduces minimal computational overhead (especially compared to running machine learning models). However, if accuracy is critical, machine learning models are more suitable.
\end{enumerate}

While our model achieves promising results, future research could explore several avenues to further enhance lock prediction in database management systems. Incorporating additional workload features into model training, such as query execution plans and transaction logs, could improve predictive accuracy by providing richer contextual information. Additionally, developing real-time adaptive models that dynamically adjust to workload shifts would enhance robustness and applicability in dynamic database environments. Another key direction is the integration of lock prediction within DBMS query optimizers, enabling real-world deployment and validation of predictive locking mechanisms. By addressing these challenges, lock prediction models could evolve into a fundamental component of next-generation intelligent database management systems, optimizing performance and resource utilization.

\section{Threats to Validity}
While our study demonstrates the effectiveness of deep learning models in predicting database locks, several potential threats to validity must be considered. We categorize these threats into internal, external, and construct validity concerns as per~\cite{yin2017case}.

\subsection{Internal Validity}
Internal validity refers to factors that may introduce bias or affect the correctness of our results. One potential threat may arise from the use of a specific IBM Db2 environment, which may introduce data collection bias and cause the models to overfit to environment-specific workload patterns. Additionally, the choices made during feature selection and encoding may fail to capture important latent variables that influence lock behavior, thereby affecting the predictive accuracy of the models. Another factor influencing internal validity is the tuning of model hyperparameters, which may unintentionally bias the models toward particular characteristics of the training data, limiting generalizability even within the same environment. While we attempt to mitigate these concerns through consistent data preprocessing and multiple training iterations using different random seeds, these limitations should be considered when interpreting the results.

\subsection{External Validity}
External validity concerns the generalizability of our findings beyond the studied environment. Our approach is tested on IBM Db2, and its applicability to other DBMS, such as MySQL or Oracle, remains to be validated. Furthermore, the scalability of our model in high-load transactional systems and its adaptability to different workload types, such as OLTP versus OLAP, require further investigation. However, the same empirical evaluation and analysis methods can be applied to other software products, provided that well-designed and controlled experiments are conducted.

\subsection{Construct Validity}
Construct validity assesses whether our study accurately measures what it intends to. Predicting locks is a proxy for performance improvements; however, additional experiments are needed to measure the real-world benefits, such as reduced transaction latency. Our success metrics focus on accuracy and sequence prediction, but alternative evaluations, such as query execution time reduction, could offer deeper insights. Additionally, the model’s effectiveness over time must be validated against evolving database workloads.

\subsection{Mitigation Strategies}
To address these threats, future research should validate our model on multiple DBMS platforms, conduct real-world deployment testing, and implement adaptive learning techniques to handle shifting transaction behaviors. These strategies will help refine our approach and improve the reliability of predictive models for database lock management.

\section{Conclusion}
In many enterprise settings, holistic database encryption provides strong protection but may require prolonged downtime and increased storage overhead. These demands make it difficult to implement online encryption in high-throughput database environments without disrupting critical operations. 
To address this challenge, we envision a solution that enables online database encryption aligned with system activity, eliminating the need for downtime, excess storage, or full-database reprocessing. Central to this vision is the ability to predict which parts of the database will be accessed next, enabling encryption to be applied incrementally and just in time. 

This study takes a step toward this solution by proposing a deep learning-based approach to predict database locks in IBM Db2 workloads, a key step in online database encryption with minimal overhead. By addressing key research questions, we demonstrated the feasibility of forecasting both the next table and data page to be locked while maintaining prediction stability over time. Our findings highlight that historical lock sequences can be leveraged to improve the accuracy of lock predictions, providing a potential pathway toward enabling online database encryption with minimal performance overhead or service disruption.

The implications of our work extend to secure and efficient database management, including adaptive encryption scheduling, contention-aware transaction processing, and workload-driven key management. By anticipating lock patterns, database systems can dynamically optimize encryption operations, ensuring data security while minimizing disruptions to ongoing transactions. Future research can explore the integration of our approach in real-time database environments to validate its impact on live encryption processes and workload. Further optimizations in model performance and computational efficiency could help bridge the gap towards optimal online database encryption features.

\begin{acks}
This work was partially supported by the Natural Sciences and Engineering Research Council of Canada (Grant \# RGPIN-2022-03886) and the IBM Centre for Advanced Studies (Project \#1121). The authors thank the Digital Research Alliance of Canada for providing computational resources.
\end{acks}

\bibliographystyle{ACM-Reference-Format}
\bibliography{refs}

\end{document}